\documentstyle[12pt,aasms4,apjfonts,psfig]{article}
\lefthead{Cusumano et al.}

\righthead{Detection of 33.8 ms X-ray pulsations in SAX~J0635$+$0533}

\begin{document}

\title{Detection of 33.8 ms X-ray pulsations in SAX~J0635$+$0533}

\author{G.~Cusumano, M.C.~Maccarone, L.~Nicastro, B.~Sacco}
\affil{Istituto di Fisica Cosmica con Applicazioni all'Informatica, CNR,\\ 
      Via U. La Malfa 153, I-90146, Palermo, Italy;
 'surname'@ifcai.pa.cnr.it}

\and

\author{P.~Kaaret}
\affil{Harvard-Smithsonian Center for Astrophysics,\\
	60 Garden St., Cambridge, MA 02138, USA; pkaaret@cfa.harvard.edu}
%
%

\begin{abstract}

We revisited the {\it Beppo}SAX observation of
SAX~J0635$+$0533, suggested as a possible counterpart to the
gamma-ray source 2EG~J0635$+$0521.  We have discovered a
33.8~ms pulsation from the source and derived an improved
position, consistent with the location of the Be star proposed
as a binary companion.  We interpret the periodicity as the
spin period of a neutron star in a binary system with a Be
companion.  

\end{abstract}

\keywords{X-rays: stars --- pulsars: general --- gamma rays:
observations --- stars: individual (2EG J0635+0521,
SAX~J0635+0533) --- stars: neutron}

\section{Introduction}

The X-ray source SAX~J0635$+$0533 was discovered by Kaaret et
al.\ (1999) thanks to a {\it Beppo}SAX observation within the
error box of the unidentified  Galactic gamma-ray source
2EG~J0635$+$0521 (\cite{thomson}), a candidate gamma-ray
pulsar as suggested by its hard gamma-ray spectrum
(\cite{merck}).  The X-ray source is characterized by quite
hard X-ray emission detected up to 40~keV (\cite{kaaret}). 
Its energy spectrum is consistent with a power-law model with
a photon index of 1.5, an absorption column density of $2.0
\times 10^{22} \rm \, cm^{-2}$, and a flux of $1.2 \times
10^{-11} \rm \, erg \, cm^{-2} \, s^{-1}$ in the $2-10$~keV
energy band.  A search for pulsed emission over a period 
range from 0.030~s to 1000~s did not detect any pulsed
signal.  Due to the large error box of the gamma-ray source,
the identification of SAX~J0635$+$0533 with 2EG~J0635$+$0521
is not definitive: such an identification could only be made
through pulsed detection in both X-ray and gamma-ray emission
or a much improved gamma-ray position.

Follow up optical observations (\cite{kaaret}) suggest as a
counterpart of SAX~J0635$+$0533 a Be star with a V-magnitude
of 12.8, located within the $1'$ X-ray source error box. The
estimated distance is in the range $2.5-5$~kpc. The total
Galactic 21~cm column density along the optical counterpart
direction is $7 \times 10^{21} \rm \, cm^{-2}$ 
(\cite{stark}), in agreement with an estimation derived from
the extinction of the optical spectrum.  Kaaret et al.\ (1999)
assert that the larger column density for SAX~J0635$+$0533,
estimated from the X-ray spectrum, implies the presence of
circumstellar gas around the X-ray source.  Moreover, taking
into account the positional consistency between the X-ray and
the gamma-ray sources as well as the hard X-spectrum, the
strong X-ray absorption, and the optical association with the
Be star, they suggest that SAX~J0635$+$0533 is an X-ray binary
emitting gamma-rays.

We revisited the {\it Beppo}SAX observation of SAX
J0635$+$0533, available from the public archive
(obs.code \#30326001).
  In this letter we present new imaging and timing
results.  Our analysis has revealed a 33.8~ms pulsation of the
X-ray source. \S 2 describes the observation and data
reduction.  In \S 3,  we report the data analysis procedures
and results.  We conclude with a discussion of the results in
\S 4.

\section{Observation and Data Reduction}

The field including the serendipitous source SAX~J0635$+$0533
was observed on 23-24 October 1997 by the Narrow Field
Instruments (NFIs) on board {\it Beppo}SAX (\cite{boella1}).
In our analysis, we use only data coming from the two NFI
imaging instruments, namely  the Low Energy Concentrator
Spectrometer (LECS) operating in the energy range 0.1--10~keV
(\cite{parmar}) and the Medium Energy Concentrator Spectrometer
(MECS) operating in the energy range 1.3--10~keV
(\cite{boella2}). At the observation epoch only two of the
three MECS detector units were operating.

LECS and MECS raw data have been reduced in cleaned photon
list files by using the SAXDAS v.2.0 software package and
adopting standard selection
criteria\footnote{http://www.sdc.asi.it/software/saxdas}.  The
source was located $\sim 3'$ off-axis. The total duration of
the observation is 72\,714~s and the net exposure is 14\,043~s
and 34\,597~s for LECS and MECS, respectively. The different
exposures are due to the fact that the LECS was only operated
during satellite nighttime.

\section{Data Analysis}

In our analysis, we include only events within a $3.7'$ radius
around the source position; this radius optimizes the
signal--to--noise ratio for SAX~J0635$+$0533  and contains
$\sim$80\% of the source signal for both LECS and MECS.  Due
to the fact that the source is strongly absorbed, we further
improve the signal--to--noise ratio by using the energy range
$1.8-10$ keV for both the instruments. The resulting number of
selected  events is then 479 for LECS and 3\,658 for MECS,
respectively. We estimate that less than $1\%$ of these events
is due to the diffuse and instrumental background.

\subsection{Spatial Analysis}

The position of SAX~J0635$+$0533 has been re-evaluated by
considering MECS data accumulated under specific constraints
of the spacecraft; in particular, only data referring to the Z
star tracker (the one aligned  with the NFIs) in use have been
taken into account.  This allows us to reduce the attitude
reconstruction error to $\sim 30''$ (90\% confidence 
level)\footnote{http://www.sdc.asi.it/software/cookbook/attitude.html}. 

After suitable smoothing of the screened data and then by
using a Gaussian  centering procedure, the position has been
found at
 $RA=06^{\rm h}35^{\rm m}18^{\rm s}$, $DEC=05^{\circ}33'11''$
(J2000)  with no significant statistical error.  The error box
is only determined by the attitude reconstruction error. 
The Be star (Kaaret et al.\ 1999) is within the $30''$
of the SAX~J0635$+$0533 error box and $6.8''$ distant from its
center. 

\placefigure{fig1}

Fig.~1 shows a finding chart for SAX~J0635$+$0533 from the 
Digitized Sky Survey in a field of $4' \times 4'$.  Within the
reduced X-ray source error circle, only two of the seven stars 
referred to in Kaaret et al.\ (1999) are now present.  The Be
star is the brightest one which is positioned close to the
center of the X-ray source error circle.

\subsection{Temporal Analysis}

The SAX~J0635$+$0533 light--curve (1000~s bin size) for the
MECS is shown in Fig.~2a (gaps are present due to
non--observing time intervals during South Atlantic Anomaly
and Earth occultation).  Fig.~2a shows that the emission of
SAX~J0635$+$0533 is variable up to a factor of 10.  Fig.~2b
shows the hardness ratio between 4--10~keV and 1.8--4~keV
bands. The relative contributions of these two energy bands
are not constant, but no strong correlation with respect to
the total intensity is evident. 

\placefigure{fig2}

In order to search for periodicity, the arrival times of all
selected events have been converted to the Solar System
Baricentric Frame, using the BARICONV code\footnote{
http://www.sdc.asi.it/software/saxdas/baryconv.html}. The
$Z_1^{2}$ test (\cite{buccheri}) on the fundamental harmonics 
with the maximum resolution ($\delta f = 1/72\,714 \rm \,$ Hz)
applied to the MECS baricentered arrival times does not reveal
significant deviations from a statistically flat distribution
up to 50~Hz.

If  SAX~J0635$+$0533 is a binary pulsar of rotational spin
period $P_s$ and orbital period $P_o$, the observed $P_s$ is
modulated by the orbital motion.  Thus, a direct search for a
coherent oscillation at $P_s$ can be successful only if the
modulation amplitude is small over the time interval $\Delta
T$ in which the search is performed.  This condition is
satisfied if $\Delta T \ll P_o$.  To reduce the effect of a
possible orbital motion in the periodicity search, we divide
the whole data span into $M$ subintervals, calculating the
$Z_1^2$ statistics for each trial period in each subinterval,
and then adding together the $M$ statistics for each trial
period.  This procedure results in a less noisy spectrum.  The
sum of the $M$ separate $Z_1^2$ statistics results in a
statistics with $2M$ degrees of freedom (\cite{bendat}) which
we refer to as the $Z^2_1 (\nu = 2M)$ statistics.  Because the
power depends on the square of the pulsed signal, its 
strength decreases with $M$ and only sufficiently strong
signals can be detected.  We selected time slices
corresponding to intervals of continuous observation taken
between two Earth occultation periods.  The total number of
these slices is $M = 13$, each one lasting $\sim 
3300 \rm \, s$.  We adopted a frequency step 
$\delta f = 2 \times 10^{-4} \rm \, Hz$, spanning 50~Hz of
search range with 250\,000 trial frequencies. 

\placefigure{fig3}

Fig.~3 shows the power spectrum obtained with the MECS and LECS data 
in the $Z^2_1 (\nu = 26)$ statistics as
a function of frequency, where an evident excess appears at 
$f_0=29.5364 \pm 0.0001$ Hz. The value of $Z^2_1 (\nu = 26)$
at this frequency is equal to $99.6$. Because the $Z^2_1 (\nu
= 26)$ follows the $\chi^2$  statistics with 26 degrees of
freedom, the single trial chance occurrence probability to
have an excess greater than 99 is $2 \times 10^{-10}$, as shown in Fig.~4.  
Taking into account the number of trial frequencies used, the
probability is $5 \times 10^{-5}$,  corresponding to 
4 standard deviations of the Gauss statistics.  
When we use only MECS data, the $Z^2_1 (\nu = 26)$ value
decreases to 91 in agreement with the reduction of the source
counts.  We tested also the power at half and double values of
the detected frequency $f_0$: no signal is present at 
$f=f_0/2$ or at $f=2f_0$.
In order to check for the persistence of the periodicity in the whole
observation, we also binned the maximum resolution spectrum of the entire 
data set by a factor 13. We obtain a high $Z^2_1 \; (\nu = 26) > 104$, at 
frequency = 29.5364 Hz, indicating that the previous high power value does 
not come from one or few selected time intervals.

\placefigure{fig4}

From the $Z^2_1 (\nu = 26)$ value we can estimate
the pulsed fraction.  For $N_p$ pulsed counts over $N_t$
total counts, $Z^2_1 (\nu = 26) = 2 \alpha N_p^2/N_t + \nu$,
where $\alpha$ is a shape constant.
In our case, $\alpha=0.25$ (sinusoidal shape), 
and the pulsed fraction is then about 0.2.

We also performed a periodicity search over the full observation including 
the first and second derivative of the frequency, $\dot{f}$ and $\ddot{f}$ 
respectively, at the maximum resolution step. The search was performed 
within the interval $\Delta f = 2 \times 10^{-4}$ Hz centered on the 
detected value $f_0$. The value range for $\dot{f}$ and $\ddot{f}$ were chosen
consequently. We obtained a maximum in the power spectrum 
($Z_1^2 (\nu =2) = 52$) for  $f = 29.53643 \pm 0.00001 $ Hz, 
$\dot{f} = (-3.1 \pm 0.2) \times 10^{-9}$ Hz s$^{-1}$ and  
$\ddot{f} = (1.1 \pm 0.1 ) \times 10^{-14}$  Hz s$^{-2}$, where the errors 
refer to the parameter resolution.  
Assuming a circular orbit, the second order polynomial expansion of the frequency vs.
time relation is locally compatible with a orbital period $P_o \sim 18$ days and semi-major 
projected axis 
$a_p \sin i \sim 63$ s lt.
However, we stress that 
the interval of observation is
much shorter than the derived orbital period, and that the parabolic
fit may not be a good representation if the orbit is eccentric.  Thus,
these results should be taken only as a possible indication of the
orbital parameters and not as a firm detection of orbital motion.\\
Figure 5 shows the pulse profile folding the data taking into account
the frequency derivative terms. Note that if the orbit is 
eccentric, the real pulse profile could be narrower.

\placefigure{fig5}

\section{Discussion}

The coherence of the detected periodicity is high, $Q =
f/\delta f \sim 10^5$.  This value is much greater than those
observed in quasiperiodic oscillations (QPOs) often detected
in the X-ray emission of X-ray binaries (\cite{vanderklis}). 
The high coherence induces us to interpret this periodicity as
a neutron star spin period.

The association between SAX~J0635$+$0533 and the Be star has been 
reinforced thanks to a reduced error circle of the X-ray position 
coordinates. Timing analysis results indicate that the neutron star  
could orbit around a companion star.
The tentative orbital parameters are consistent, for orbit inclination
less than $25^{\circ}$, with a primary mass greater than 10 $M_\odot$, 
as expected for a Be-star.

The X-ray emission may be powered either by accretion or by
spin-down of the neutron star.  We consider these
possibilities in turn.

The SAX~J0635$+$0533 system may consist of a rotation-powered
pulsar orbiting the Be star.  In this case, the X-ray emission
could be magnetospheric emission similar to the power-law
component in the X-ray emission of known X-ray/gamma-ray
pulsars.  The pulsation frequency we have found and the X-ray
luminosity of the source are similar to those of the known
X-ray/gamma-ray pulsars.  The high X-ray variability of
SAX~J0635$+$0533 on time scales of 1000~s is unlike the steady
X-ray emission seen from isolated pulsars.  However, it could
be produced by variable X-ray absorption caused by matter in
the binary system or a wind from the Be star.

An alternative, but still rotation-powered, scenario is that 
SAX~J0635$+$0533 is similar to the Be radio pulsar
PSR~J1259--63.  These two sources have similar spin
frequencies and similar X-ray spectra (\cite{nicastro}).  In
this case, the X-ray emission of SAX~J0635$+$0533 should arise
from a  shock interaction of the energetic particles from the
pulsar with the wind from the Be star.  However, the upper
bound, 8\%, on the X-ray pulsed fraction from PSR~J1259--63 in
the 2--10~keV band (Kaspi et al.\ 1995) is well below the
value estimate for SAX~J0635$+$0533 in the same energy band.

The X-ray emission from SAX~J0635$+$0533 may be powered by
accretion.  Strong X-ray variability would naturally occur in
such a system.  Following this interpretation we can infer the
magnetic field strength of the neutron star.  For accretion to
proceed, the centrifugal force on the accreting matter
co-rotating in the magnetosphere must be less than the local
gravitational force (\cite{illarionov}; \cite{stella}). 
Assuming a bolometric luminosity of $1.2 \times 10^{35} \rm \,
erg \, s^{-1}$ (0.1--40 keV) estimated from  spectra results
given in Kaaret et al.\ (1999) for a 5 kpc distance, and a
neutron star mass of $1.4 \rm \, M_{\odot}$ and radius of
10~km, we can set an upper limit on the magnetic field
strength of $2 \times 10^{9} \rm \, G$.  This is a factor
$10^3$ lower than those measured in typical accreting X-ray
pulsars, but similar to the fields inferred for the 2.49~ms
low-mass X-ray binary SAX~J1808.4$-$3658 (\cite{wijnnands})
and for millisecond radio pulsars.  The X-ray luminosity of
SAX~J0635$+$0533 is a factor of 10 below that of most Be/X-ray
binaries or the peak luminosity of SAX~J1808.4$-$3658, but may
simply indicate a low mass accretion rate.

A definitive association of SAX~J0635+0533 with the EGRET
source requires detection of a periodicity in gamma-rays at
the pulsar spin period.  Due to the long integration time
required to obtain a detectable gamma-ray signal, only a priori
knowledge of the binary parameters would permit a sensitive
search for periodicity in gamma-rays.  This can  be obtained
with additional X-ray observations of SAX~J0635$+$0533.

\acknowledgments
We wish to thank Enrico Massaro (University of Rome) and 
Ignacio Neguerela ({\it Beppo}SAX SDC) for scientific discussions, and
Guido Vizzini (IFCAI) for the technical support in the data reduction 
procedure.


\clearpage

\figcaption[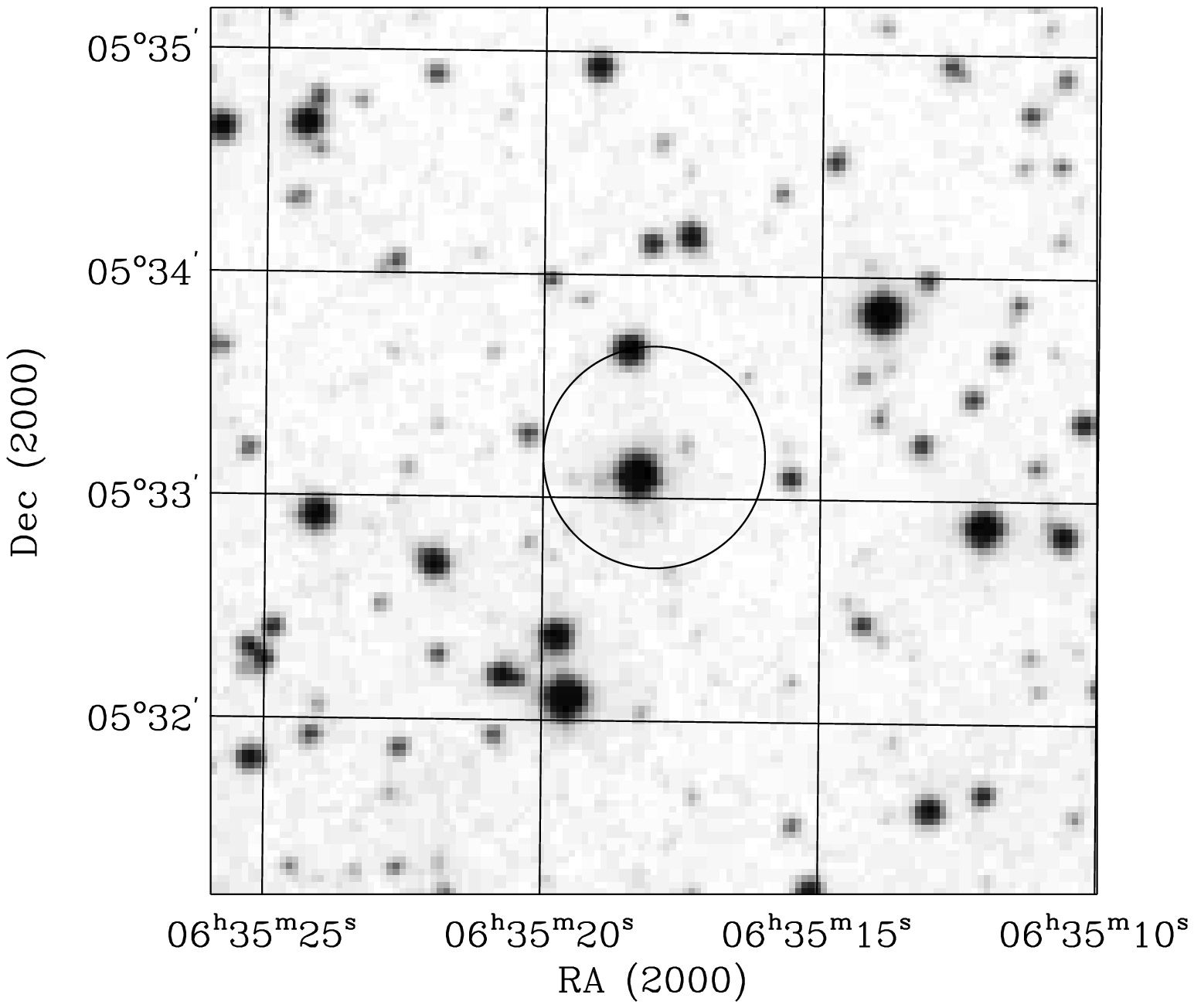]{Finding chart for SAX~J0635$+$0533 from the 
Digitized Sky Survey in a field of $4' \times 4'$. 
The X-ray error circle has a radius of $30''$ (see text). \label{fig1}}

\figcaption[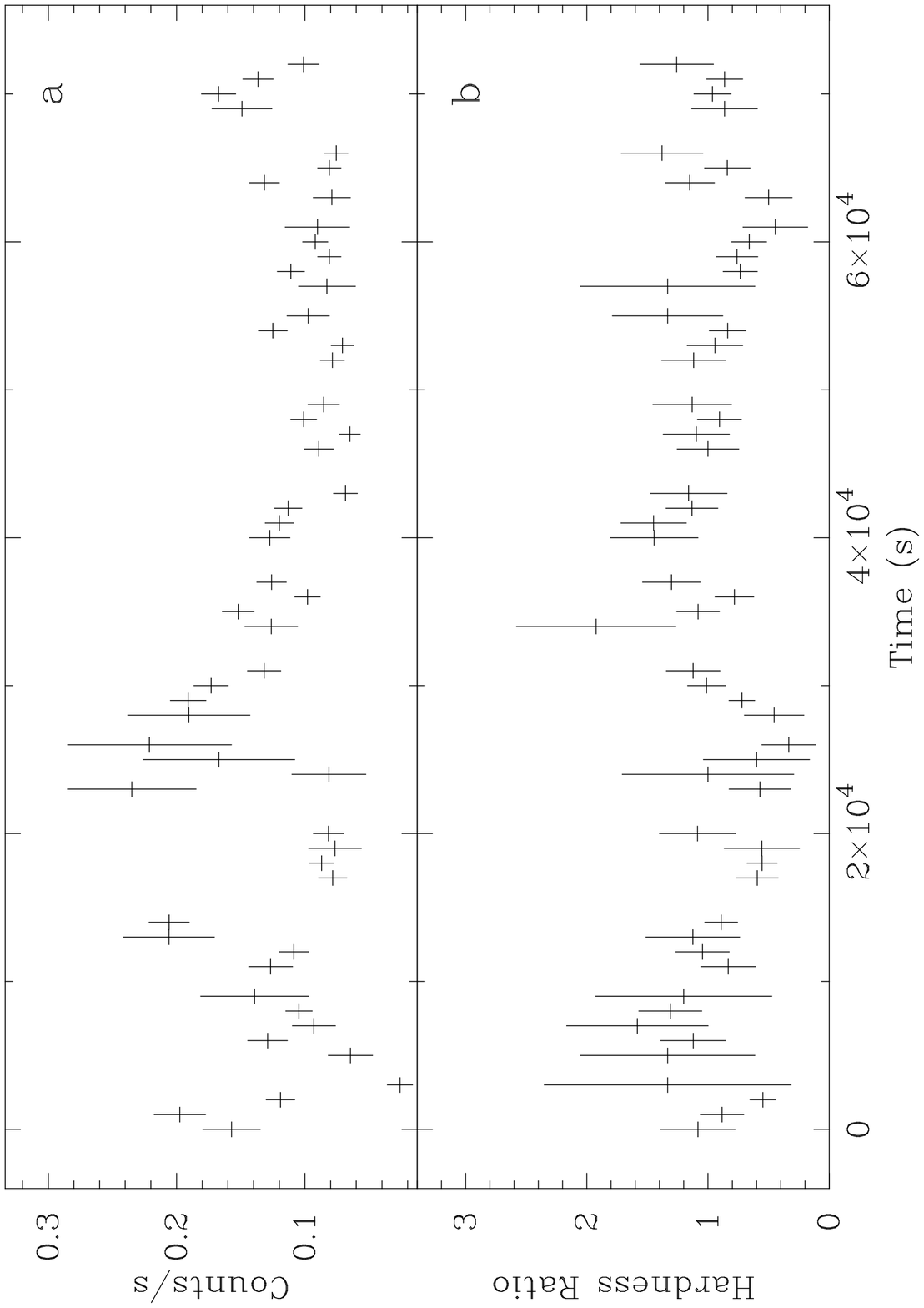]{a) The MECS light curve of SAX~J0635$+$0533. 
b) Hardness ratio vs time between $4-10$ and $1.8-4$ keV energy 
bands. \label{fig2}}

\figcaption[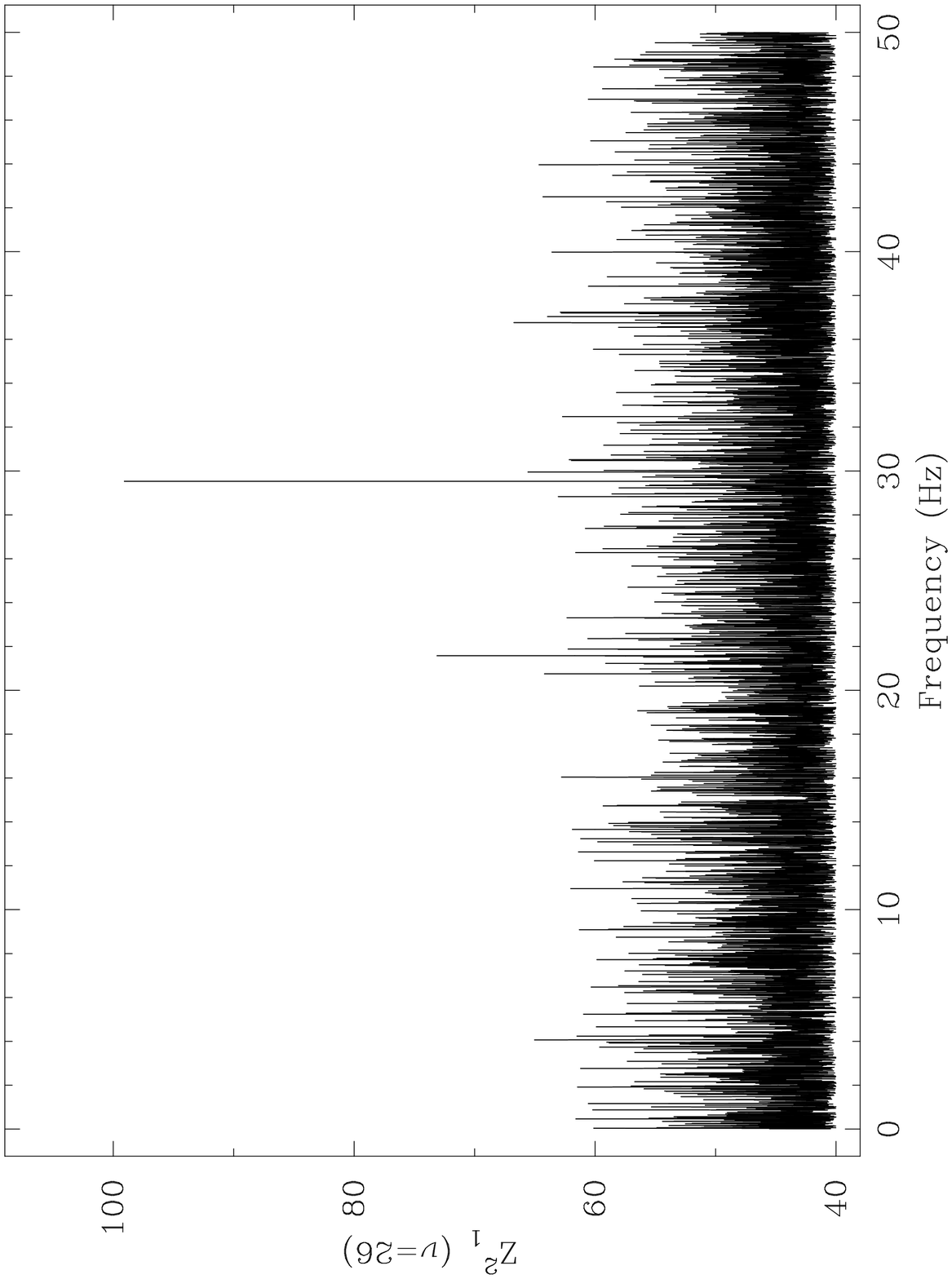]{The $Z^2_1 (\nu = 26)$ plot for MECS+LECS 
events. \label{fig3}}

\figcaption[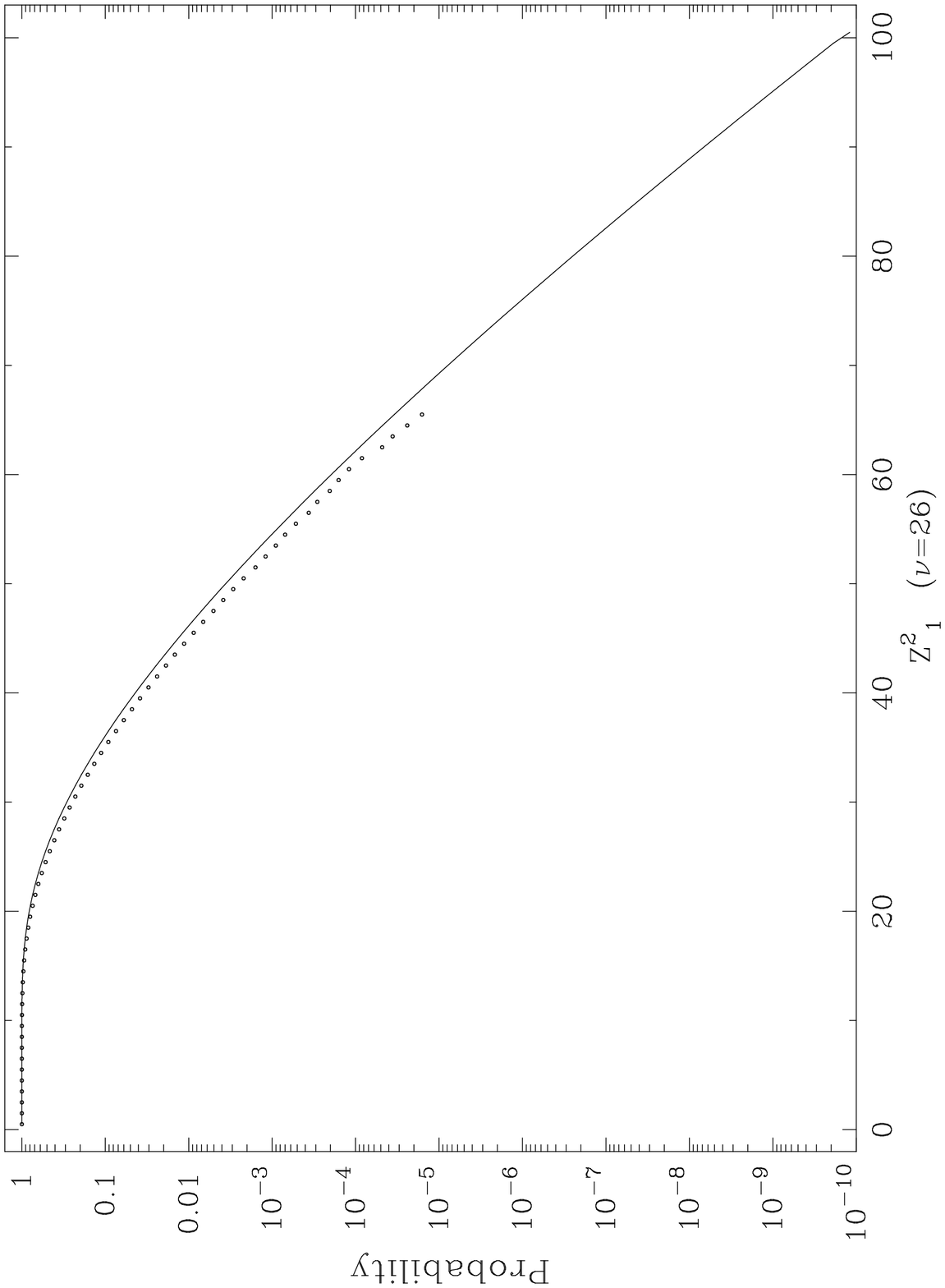]{The chance occurrence probability to find 
values greater than $Z^2_1 (\nu = 26)$ (continuous line) compared with the
measured integral distribution (dots). \label{fig4}}

\figcaption[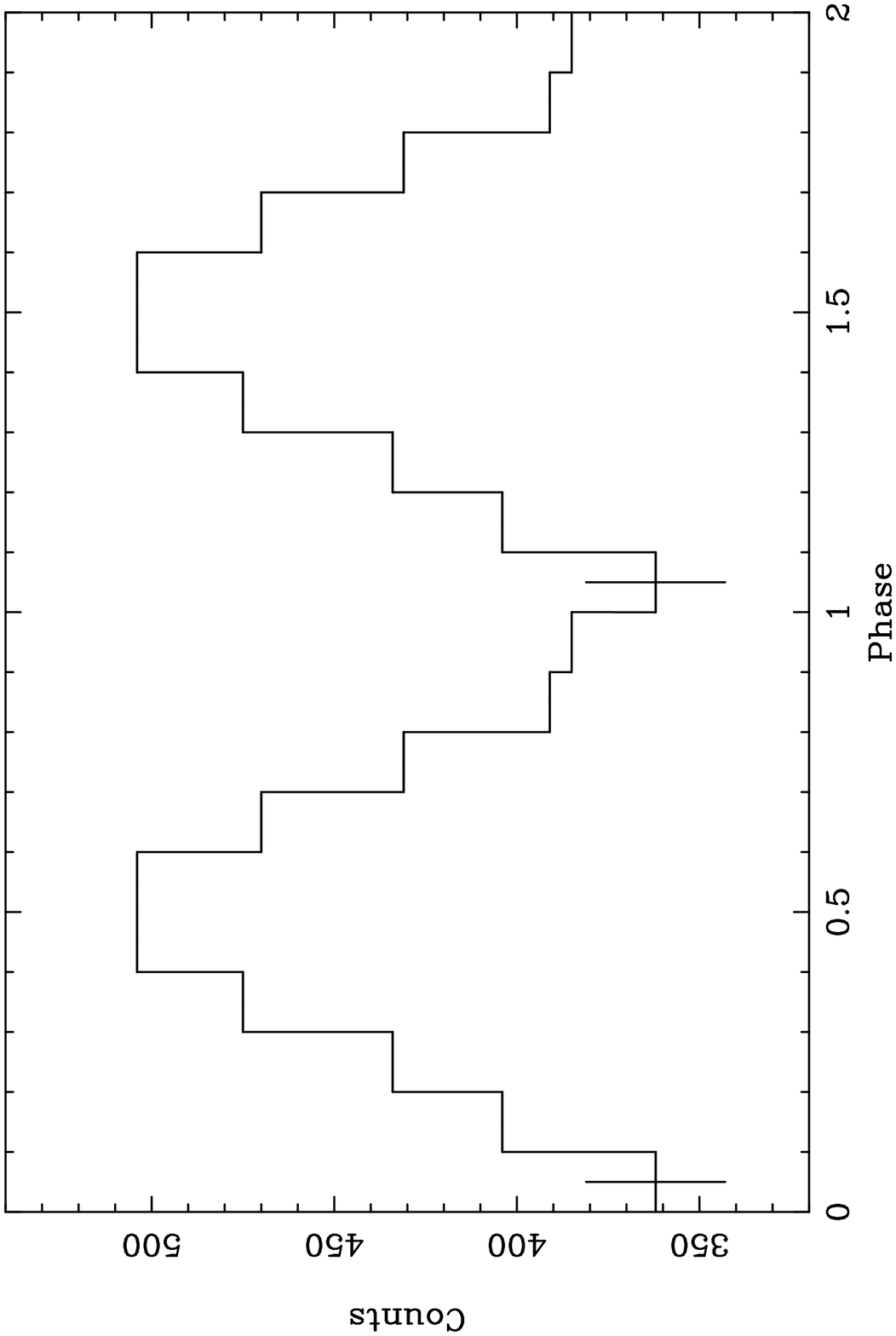]{Pulse profile obtained taking into account
 $\dot{f}$ and $\ddot{f}$. Two cycles are shown for clarity.
The superimposed error bar correspond to 1 standard deviation. \label{fig5}}
\end{document}